\documentclass[letterpaper]{article} 
\usepackage{aaai2026}  
\usepackage{times}  
\usepackage{helvet}  
\usepackage{courier}  
\usepackage[hyphens]{url}  
\usepackage{graphicx} 
\usepackage{natbib}  
\usepackage{caption} 
\usepackage{amsmath}
\usepackage{amsfonts}
\usepackage{threeparttable}
\usepackage{multirow}
\usepackage{booktabs}
\usepackage{float}
\usepackage{enumitem}
\usepackage{amssymb}
\usepackage{pifont}
\usepackage[table,xcdraw]{xcolor}
\usepackage[ruled,vlined]{algorithm2e}

\SetCommentSty{mycommfont}

\SetKwInput{KwInput}{Input}                
\SetKwInput{KwOutput}{Output}              

\usepackage{newfloat}
\usepackage{listings}
\DeclareCaptionStyle{ruled}{labelfont=normalfont,labelsep=colon,strut=off} 
\floatstyle{ruled}
\newfloat{listing}{tb}{lst}{}
\floatname{listing}{Listing}
\title{Diff-V2M: A Hierarchical Conditional Diffusion Model with Explicit Rhythmic Modeling for Video-to-Music Generation}
\author{
    Shulei Ji\textsuperscript{\rm 1,\rm 2},
    Zihao Wang\textsuperscript{\rm 1,\rm 3},
    Jiaxing Yu\textsuperscript{\rm 1},
    Xiangyuan Yang\textsuperscript{\rm 4},\\
    Shuyu Li\textsuperscript{\rm 1},
    Songruoyao Wu\textsuperscript{\rm 1},
    Kejun Zhang\textsuperscript{\rm 1,\rm 2}\thanks{Corresponding author.}
}
\affiliations{
    \textsuperscript{\rm 1}Zhejiang University\\
    \textsuperscript{\rm 2}Innovation Center of Yangtze River Delta, Zhejiang University\\
    \textsuperscript{\rm 3}Carnegie Mellon University\\
    \textsuperscript{\rm 4}Xi'an Jiaotong University\\
    \{shuleiji, carlwang, yujx\}@zju.edu.cn, ouyang\_xy@stu.xjtu.edu.cn, \{lsyxary, wsry, zhangkejun\}@zju.edu.cn
}

\begin{document}

\maketitle

\begin{abstract}
Video-to-music (V2M) generation aims to create music that aligns with visual content. However, two main challenges persist in existing methods: (1) the lack of explicit rhythm modeling hinders audiovisual temporal alignments; (2) effectively integrating various visual features to condition music generation remains non-trivial. To address these issues, we propose Diff-V2M, a general V2M framework based on a hierarchical conditional diffusion model, comprising two core components: visual feature extraction and conditional music generation. For rhythm modeling, we begin by evaluating several rhythmic representations, including low-resolution mel-spectrograms, tempograms, and onset detection functions (ODF), and devise a rhythmic predictor to infer them directly from videos. To ensure contextual and affective coherence, we also extract semantic and emotional features. All features are incorporated into the generator via a hierarchical cross-attention mechanism, where emotional features shape the affective tone via the first layer, while semantic and rhythmic features are fused in the second cross-attention layer. To enhance feature integration, we introduce timestep-aware fusion strategies, including feature-wise linear modulation (FiLM) and weighted fusion, allowing the model to adaptively balance semantic and rhythmic cues throughout the diffusion process. Extensive experiments identify low-resolution ODF as a more effective signal for modeling musical rhythm and demonstrate that Diff-V2M outperforms existing models on both in-domain and out-of-domain datasets, achieving state-of-the-art performance in terms of objective metrics and subjective comparisons. Demo and code are available at \url{https://Tayjsl97.github.io/Diff-V2M-Demo/}.

\end{abstract}


\section{Introduction}
Music not only activates the auditory system but also modulates visual perception through its functional connections with the visual cortex \cite{0}. As such, background music serves as a critical element in enhancing the overall impact and expressiveness of videos. However, traditional background music composition often relies on manual editing or customized production, which is both costly and inflexible. With the rapid rise of video streaming platforms like YouTube and TikTok, alongside the emergence of video generative models such as Sora \cite{53} and Veo \cite{54}, the demand for personalized audiovisual content has surged. In this context, video-to-music generation has emerged as a rapidly growing research topic.

In recent years, video-to-music generation has attracted increasing research attention, enabling background music creation tailored to diverse video domains \cite{00,55}. Early works primarily targeted \textbf{human-centric videos}, such as silent instrument performances and dance clips. Studies on instrument performance videos \cite{1,2,3} generated music by modeling visual cues from performers, while works on dance videos \cite{4,5,6} usually extracted human motion features to guide music generation.
Moving beyond human-centric videos, more recent research \cite{8,9,10,11} expanded to \textbf{general videos} such as music videos and movie trailers by extracting multi-perspective visual features and incorporating multi-condition guidance to steer music generation. In parallel, some studies \cite{12,13} incorporated video understanding into large language models (LLMs), translating videos into textual prompts that condition text-to-music generation pipelines.

Despite recent advances, existing video-to-music generation methods lack explicit modeling of musical rhythm, which is crucial for achieving precise audiovisual temporal alignment. Existing approaches model visual dynamics through scene detection \cite{15}, optical flow \cite{8}, frame differences \cite{14,15}, or frame-level visual features \cite{9,10}. However, these strategies still require the model to implicitly learn the mapping from visual dynamics to musical rhythm. Others \cite{12,13} attempt to translate videos into textual prompts, which often fail to preserve fine-grained temporal dynamics. There remains a lack of a unified and effective musical rhythmic representation that can support consistent temporal alignment in general video-to-music generation.

The other key challenge lies in how to effectively integrate diverse visual features to guide music generation. The fusion of multi-perspective features from videos, such as emotional, semantic, and rhythmic features, remains non-trivial. In prior studies, progressive fusion strategies \cite{14,16} often involve multi-stage architectures that increase computational overhead, while simple concatenation~\cite{15,17} fails to capture the underlying dependencies between features. Alternatively, large language models (LLMs) have been leveraged to convert video content into text-based conditions, thereby simplifying inputs and bypassing explicit feature fusion. However, textual descriptions struggle to capture dynamic visual cues, limiting the temporal alignment between video and generated music.

To address the aforementioned challenges, we propose \textbf{Diff-V2M}, a hierarchical conditional diffusion transformer framework designed for general video-to-music generation. Inspired by TiVA \cite{19}, which uses low-resolution mel-spectrograms as audio layouts to support temporal synchronization, we systematically explore and compare several rhythmic representations, including low-resolution mel-spectrograms, tempograms, and onset detection functions (ODF) \cite{20}. To ensure robust rhythm conditioning, a rhythm predictor is trained to infer rhythmic representations from video and is jointly optimized with the music generator during training following the proposed scheduled conditioning training strategies. In addition to rhythmic features, Diff-V2M extracts color histograms \cite{14,21} as emotional features and CLIP features \cite{22} as semantic cues, enabling emotional and semantic alignment between video and music. Moreover, to condition the generator on these three features, we design a hierarchical conditional module. Specifically, emotional features are first integrated through a cross-attention layer to guide the overall affective tone. Subsequently, semantic and rhythmic features are processed independently via parallel cross-attention and adaptively fused using a set of timestep-aware fusion strategies, including feature-wise linear modulation (FiLM) \cite{18} and weighted fusion.

The main contributions of this paper are as follows:
\begin{itemize}
    \item We introduce three rhythmic representations to model temporal alignment for video-to-music generation and identify low-resolution ODF as the most effective.
    \item We propose Diff-V2M, a conditional diffusion transformer framework tailored for the general video-to-music generation task. It integrates emotional, semantic, and rhythmic features via hierarchical cross-attention, enhanced by timestep-aware FiLM and weighted fusion strategies for effective multi-feature conditioning. 
    \item Extensive experiments demonstrate that Diff-V2M outperforms the state-of-the-art models in terms of objective and subjective evaluation on both in-domain and out-of-domain datasets.
\end{itemize}
\begin{figure*}[t]
\centering
\includegraphics[width=0.97\linewidth]{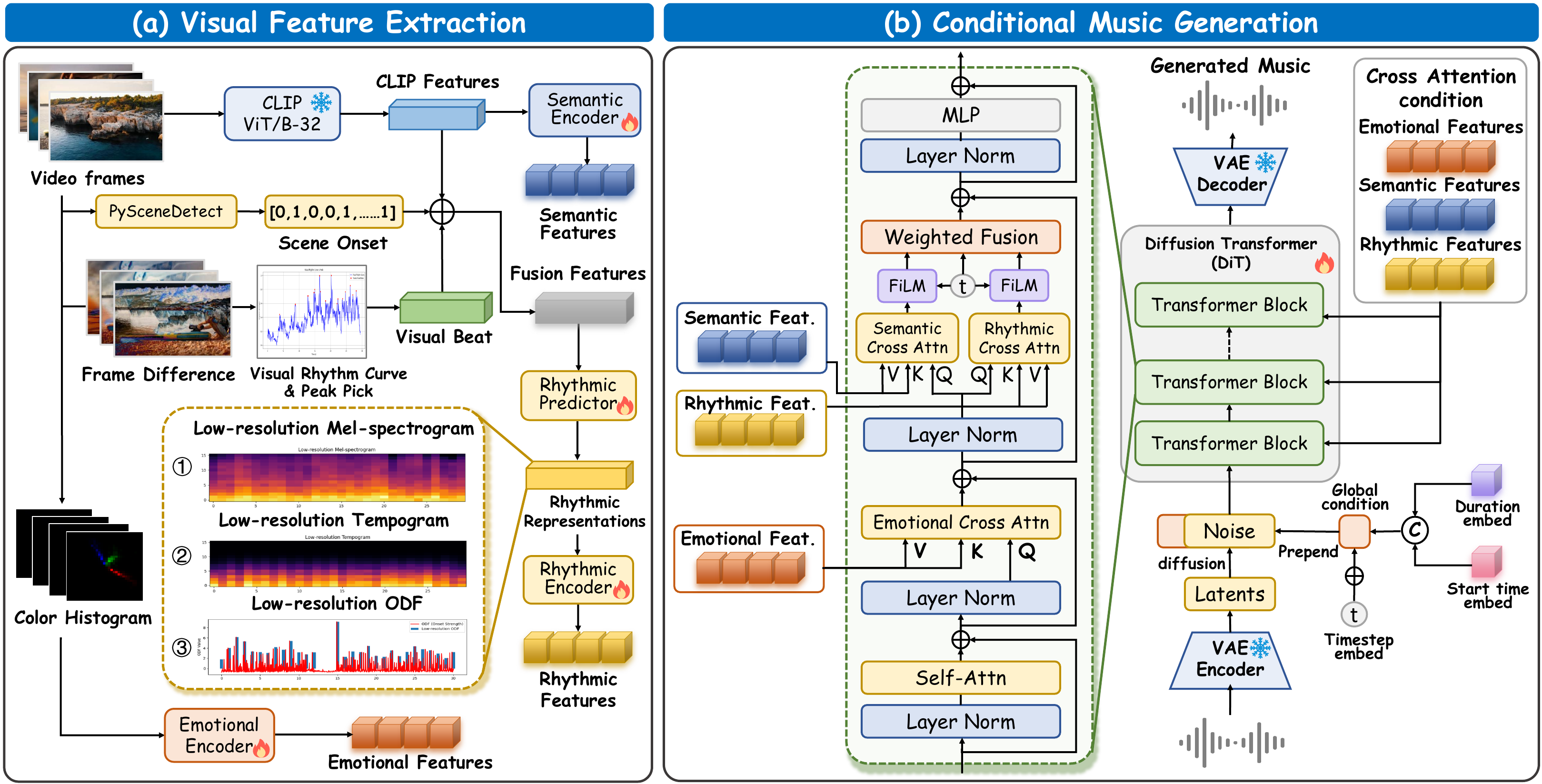} 
\caption{The architecture of Diff-V2M, consisting of two core modules: (a) visual feature extraction that derives emotional, semantic, and rhythmic features; and (b) conditional music generation built on a DiT-based LDM, which integrates multi-view features via hierarchical cross attention and timestep-aware fusion strategies.}
\label{fig1}
\end{figure*}
\section{Related Work}
\subsection{Visual Understanding}
Video-to-music generation leverages video understanding models to extract diverse visual features, including emotional, semantic, and rhythmic features. Semantic features are typically obtained using pretrained models such as CLIP \cite{22}, VideoCLIP \cite{23}, ViViT \cite{24}, and VideoMAE \cite{25}. Rhythmic features are derived from motion dynamics using models like GCN \cite{3,16}, TCN \cite{26}, and I3D \cite{5,6,27}, as well as handcrafted approaches based on optical flow \cite{8} and frame differences \cite{14,15}. Emotional features are commonly represented by frame-level color histograms \cite{14} or CLIP-based emotion probability distribution vectors \cite{15}.
In this paper, we use CLIP to extract frame-wise semantic features and model visual emotion using color histograms. To explicitly model musical rhythm, we introduce a rhythm predictor that estimates rhythmic representations directly from video.
\subsection{Music Generation}
Music generation can be broadly categorized into symbolic- and audio-domain approaches \cite{52,28}. For symbolic music generation, models such as Transformers, Variational Autoencoders (VAEs), and Generative Adversarial Networks (GANs), along with their variants, have been widely adopted \cite{29}. As music generation has evolved from unimodal to cross-modal tasks \cite{30}, such as text-to-music and vision-to-music, audio generation has gained increasing popularity due to its enhanced expressive capacity and the relative ease of collecting large-scale datasets. Autoregressive models like MusicLM \cite{31} and MusicGen \cite{32}, as well as latent diffusion models (LDMs) such as AudioLDM \cite{33} and Stable Audio \cite{34}, have achieved notable success in text-to-music generation. These models provide a strong foundation for audio-based cross-modal music generation. In this paper, we adopt an audio LDM as the backbone and advance it with a hierarchical conditioning mechanism that incorporates emotional, semantic, and rhythmic features from videos.
\subsection{Video-to-Music Generation}
Video-to-music generation can be broadly categorized by video type into human-centric videos (e.g., dance videos) and general videos (e.g., movie clips). Early studies on human-centric videos focused on silent music performance \cite{1,2,3}, while recent studies have extensively explored dance-to-music generation by leveraging motion or keypoint-based features to control rhythm and style \cite{4,5,6,7,16,17,35,36}. 
For general videos, methods typically extract diverse visual features(e.g., emotional, semantic, and rhythmic features) to guide music generation \cite{8,9,14,15,37,38}. Additionally, some approaches employ textual prompts or video captions as extra high-level control signals \cite{9,37}, while others use large language models (LLMs) to convert visual inputs into textual prompts for text-to-music generation \cite{12,13,39}. Despite these advances, existing approaches lack explicit modeling of musical rhythm and effective conditioning mechanisms for multiple visual features. Consequently, we propose a novel framework capable of generating music for diverse general videos by predicting generalizable rhythmic representations and integrating multiple video-driven features through a hierarchical conditioning module.

\section{Methodology}
\subsection{Generalizable Rhythmic Representations} \label{rhy}
Low-resolution mel-spectrograms have been proven effective for temporal control in video-to-sound effect generation \cite{19}. Motivated by this, we explore their effectiveness in video-to-music generation for the first time. In addition, we investigate tempograms and onset detection functions (ODF) as alternative rhythmic representations. To facilitate learning and improving efficiency, all representations are dimensionally reduced, as detailed below. An illustration of the three types is provided in Figure \ref{fig1}(a).

\subsubsection{Low-resolution Mel-spectrogram} are an effective control signal for coarse-to-fine audio generation \cite{19}. Given a raw Mel-spectrogram $\text{Mel}_{raw}$ of size $[M_{\text{raw}}, C_{\text{raw}}]$, we normalize and downsample it to a low-resolution version $\text{Mel}_{LR}$ with target resolution $[M, C]$:
\begin{equation}
\text{Mel}_{LR} = \text{Resize}(\text{Norm}(\text{Mel}_{raw}); M, C)
\label{eq:audio_layout}
\end{equation}
where $M_{\text{raw}}$ and $C_{\text{raw}}$ denote the original number of frames and frequency dimensions, and $M$ and $C$ are their reduced counterparts.
\subsubsection{Low-resolution Tempogram.} A tempogram is a time–tempo representation that captures the local tempo of an audio signal as it evolves over time. Following the same strategy as for $\text{Mel}_{LR}$, we normalize and downsample the raw tempogram $\text{Tem}_{raw} \in \mathbb{R}^{M_{\text{raw}} \times B_{\text{raw}}}$ to obtain a compact form $\text{Tem}_{LR} \in \mathbb{R}^{M \times B}$:
\begin{equation}
\text{Tem}_{LR} = \text{Resize}(\text{Norm}(\text{Tem}_{raw}); M, B)
\label{eq:tempogram}
\end{equation}
where $M_{\text{raw}}$ and $B_{\text{raw}}$ denote the original number of frames and tempo bins, and $M$ and $B$ are their reduced counterparts. This compact representation preserves the overall tempo contour while simplifying model learning.
\subsubsection{Low-resolution ODF.} The onset detection function (ODF) converts audio into a one-dimensional time series that reflects the likelihood or intensity of note onsets over time. Compared to mel-spectrograms and tempograms, ODF provides cleaner rhythmic cues by emphasizing critical rhythmic events. Given a raw ODF curve $\mathbf{o} = [o_1, o_2, \ldots, o_T]$, where $T$ is the number of audio frames, we apply peak detection to identify onset peaks $\mathcal{P} = \{(t_i, o_{t_i})\}_{i=1}^N$, where each $t_i$ is the time (in seconds) of a detected peak and $o_{t_i}$ is the corresponding onset strength. We then map each detected peak to its nearest second and construct a second-level vector as the low-resolution ODF, i.e., $\text{ODF}_{LR} = [o_1, o_2, \ldots, o_M]$, where $M$ is the total number of seconds. For each second, we keep  $o_m$ the maximum onset strength if any peak exists, otherwise set $o_m = 0$. 

\subsection{Architecture of Diff-V2M}
As illustrated in Figure \ref{fig1}, \textbf{Diff-V2M} consists of two main modules: visual feature extraction and conditional music generation. The feature extraction module includes a \textbf{semantic encoder}, a \textbf{rhythmic encoder}, an \textbf{emotional encoder}, and a \textbf{rhythmic predictor}. The generation module features an LDM-based \textbf{music generator}, which employs hierarchical cross-attention and feature fusion mechanisms for multi-feature conditioning. Each component is detailed in the follow-up.
\subsubsection{Visual Feature Extraction and Encoding.}
Following prior work \cite{14}, we adopt frame-wise color histograms \cite{21} to capture the underlying emotion of videos. Frame-level semantic features are obtained using a pretrained CLIP model \cite{22}, while rhythmic features are obtained from one of the three representations introduced in Section \ref{rhy}. To match the input dimensions of the generator’s conditioning module, each feature is projected through a dedicated encoder composed of linear layers.
\begin{figure}[t]
\centering
\includegraphics[width=0.97\linewidth]{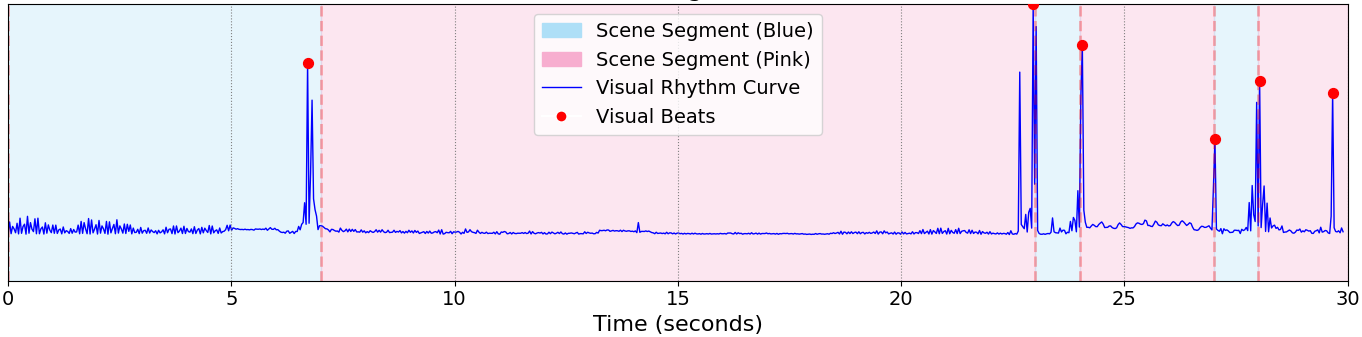} 
\caption{An example illustrating explicit video scene transitions, the visual rhythm curve, and the visual beats.}
\label{fig2}
\end{figure}
\subsubsection{Rhythmic Predictor.} 
To infer rhythmic features without relying on audio at inference, we introduce a decoder-only transformer as the rhythmic predictor that takes as input: (i) CLIP features, (ii) scene transition embeddings, and (iii) visual beat vectors.

To capture macro-level visual changes, scene transitions are detected via PySceneDetect \cite{41}, yielding a binary vector $\mathbf{e} = [e_1, e_2, \ldots, e_M]\in \{0, 1\}^M$ that marks scene boundaries per second, where $M$ is the video length in seconds and $e_m = 1$ indicates the start of a new scene at second $m$, and $e_m = 0$ otherwise. For fine-grained visual dynamics, frame-wise differences are aggregated over time to form a visual rhythm curve. Peaks are detected to obtain a second-level visual beat vector $\mathbf{v} = [v_1, v_2, \ldots, v_M] \in \mathbb{R}^M$, where $v_m$ denotes the peak beat intensity around the second $m$, or zero if no peak is detected. Figure \ref{fig2} provides an example illustrating explicit video scene transitions, the visual rhythm curve, and the visual beats selected based on peak detection. These two vectors provide complementary rhythmic cues. Although differing in granularity, they often correlate in high-activity scenes. 

To align dimensions, $\mathbf{e}$ is passed through an embedding layer $\text{Embed}(\cdot)$ and $\mathbf{v}$ through a linear projection $\text{Linear}(\cdot)$. The resulting vectors are summed with the frame-level CLIP features $C_\mathbf{s}$ to form the input sequence:
\begin{equation}
\mathbf{X} = C_\mathbf{s} + \text{Embed}(\mathbf{e}) + \text{Linear}(\mathbf{v})
\end{equation}
The resulting sequence $\mathbf{X}$ is fed into the rhythmic predictor to
estimate the target rhythmic representations introduced in Section~\ref{rhy}, enabling audio-free rhythm prediction from visual input at inference time.
\subsubsection{DiT-based Conditional Music Generator.}
We adapt Stable Audio Open \cite{34}, an LDM-based audio generator, for video-to-music generation. A VAE encodes raw waveforms into latent representations \(\mathbf{z}_a\), enabling efficient generation. A conditional diffusion model \(G\) is trained to predict the added noise $\epsilon$ from the noisy latents $\mathbf{z}_a^t$, conditions \(\mathbf{C}\), and diffusion timestep \(t\):
\begin{equation}
\mathcal{L}_{\text{LDM}} = \mathbb{E}_{t, \mathbf{z}_a^0, \boldsymbol{\epsilon}} \left[ \left\| \boldsymbol{\epsilon} - G(\mathbf{z}_a^t, \mathbf{C}, t) \right\|_2^2 \right],
\end{equation}
In this paper, condition \(\mathbf{C}\) includes emotional, semantic, and rhythmic features extracted from video, along with global embeddings (i.e., music start time and duration). A diffusion transformer equipped with a hierarchical cross-attention module is devised for integrating multiple video-driven features. 
\begin{figure*}[t]
\centering
\includegraphics[width=0.97\linewidth]{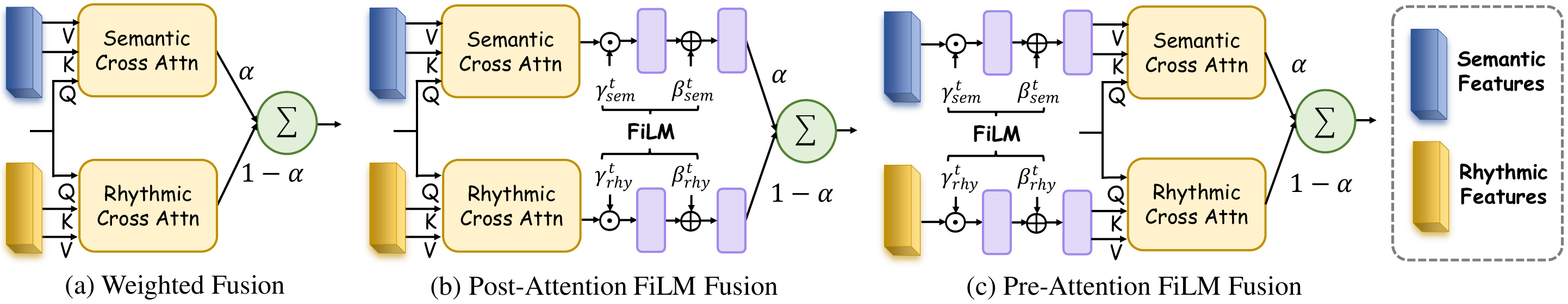} 
\caption{The illustration of different fusion strategies for semantic and rhythmic features.}
\label{fig3}
\end{figure*}
\paragraph{Hierarchical Cross-attention Module.}
As illustrated in Figure \ref{fig1}(b), the hierarchical cross-attention module first incorporates emotional features to shape the overall mood of the generated music. Next, semantic and rhythmic features are attended to via parallel cross-attention layers, preventing information entanglement and enabling more precise capture of content-relevant and tempo-aligned cues. Besides, timestep-aware feature fusion strategies are adopted to adaptively combine semantic and rhythmic branches, which will be elaborated later. This hierarchical design enables flexible integration of emotional tone, semantic meaning, and rhythmic structure for enhanced music generation.
\paragraph{Feature Fusion Strategy.}
Prior study \cite{37} proposes a feature selector that enforces exclusive attention to either semantic or dynamic features at each timestep, disregarding complementary information from other features. This rigid selection limits the model’s ability to leverage multiple features in complex video scenarios, especially when both semantic and rhythmic information are crucial. To overcome this, we introduce novel fusion strategies that adaptively balance semantic and rhythmic contributions for improved multi-feature integration. Two fusion methods are designed: weighted fusion and FiLM-based fusion.

\textbf{(1) Weighted fusion.} A gating network conditioned on the diffusion timestep \( t \) outputs a scalar weight \( \alpha \in [0,1] \) to balance semantic features \( \mathbf{h}_{\mathrm{sem}} \) and rhythmic features \( \mathbf{h}_{\mathrm{rhy}} \) . The fused feature is computed as:
\begin{gather}
\alpha = \sigma \big( f_{\mathrm{gate}}(t) \big), \
\mathbf{h}_\text{fuse} = \alpha \cdot \mathbf{h}_{\mathrm{sem}} + \big(1 - \alpha\big) \cdot \mathbf{h}_{\mathrm{rhy}}
\label{wf}
\end{gather}
where \( f_{\mathrm{gate}}(\cdot) \) denotes the gating network and \(\sigma(\cdot)\) is the sigmoid function.  

\textbf{(2) FiLM-based fusion.} To enhance fine-grained feature modulation beyond weighted fusion, we propose a Feature-wise Linear Modulation (FiLM)-based fusion mechanism. FiLM applies learnable, timestep-dependent scaling and shifting to each feature dimension, allowing precise and dimension-wise adjustment over semantic and rhythmic features. For each input feature \( \mathbf{h} \in \mathbb{R}^{B \times T \times D} \), two MLP networks generate timestep-aware modulation parameters \( \gamma^t, \beta^t \in \mathbb{R}^{B \times 1 \times D} \), which are then applied as follows:
\begin{equation}
\begin{aligned}
\text{FiLM}_\text{sem}(\mathbf{h}_{\mathrm{sem}}) &= \gamma_\text{sem}^t \cdot \mathbf{h}_{\mathrm{sem}} + \beta_\text{sem}^t \\
\text{FiLM}_\text{rhy}(\mathbf{h}_{\mathrm{rhy}}) &= \gamma_\text{rhy}^t \cdot \mathbf{h}_{\mathrm{rhy}} + \beta_\text{rhy}^t
\end{aligned}
\end{equation}

Furthermore, we investigate the optimal position for applying the above fusion strategy, as shown in Figure \ref{fig3}. Three design variants are explored:
\begin{itemize}
    \item \textbf{Weighted fusion (Figure \ref{fig3}(a))}: Semantic and rhythmic attention outputs are first computed independently and then combined via a timestep-aware weighted fusion (Eq. (\ref{wf})). When $\alpha=0.5$, this fusion becomes \textit{additive fusion} of two features. When $\alpha=1$ or 0, the fusion degrades to single \textit{feature selection} \cite{37}. 
    \item \textbf{Post-attention FiLM fusion (Figure \ref{fig3}(b))}: Each attention output is individually modulated via FiLM, followed by weighted fusion.
    
    \item \textbf{Pre-attention FiLM fusion (Figure \ref{fig3}(c))}: FiLM is applied to the semantic and rhythmic features prior to parallel cross-attention, and the resulting attention outputs are then combined via weighted fusion.
\end{itemize}
This combination of hierarchical conditioning module and feature fusion design facilitates flexible and fine-grained interaction between features, allowing the model to effectively leverage complementary cues in diverse video-to-music generation scenarios.
\subsection{Training with Scheduled Conditioning}
To mitigate the training-inference discrepancy caused by using ground-truth rhythmic representations \( C_\mathbf{r}^{\text{gt}} \) during training and predicted representations \( C_\mathbf{r}^{\text{pred}} \) during inference, we adopt a scheduled conditioning strategy that gradually substitutes ground-truth rhythmic representations with predicted ones during training.
Specifically, we define a probability schedule \( p_{\text{pred}}(e) \in [0, 1] \) to control the use of predicted rhythmic representations at epoch \( e \):
\begin{equation}
p_{\text{pred}}(e) = 
\begin{cases}
0, & \text{if } e < e_1 \\
\frac{e - e_1}{e_2 - e_1}, & \text{if } e_1 \leq e < e_2 \\
1, & \text{if } e \geq e_2
\end{cases}
\end{equation}
where \( e_1 = 10 \) and \( e_2 = 30 \) in our setup. At each epoch \( e \), a Bernoulli variable \( q\sim \text{Bernoulli}(p_{pred}(e)) \) determines whether to use predicted or ground-truth rhythmic representation. This ensures a smooth transition from teacher-forced training to fully relying on the predicted rhythms, ensuring robustness at inference time when only predicted features are available. This training strategy is inspired by \textit{Scheduled Sampling} \cite{40}, but differs in that the replaced variable is a conditioning signal rather than an autoregressive input. Last but not least, the rhythmic predictor is trained jointly with the generator, ensuring co-adaptation and better alignment with generation objectives.
\section{Experiments}
\subsection{Datasets}
We employ BGM909 \cite{37} and SymMV \cite{14} datasets for training our models. BGM909 is built upon the POP909 dataset \cite{46}, which includes 909 piano arrangements of Chinese pop songs accompanied by temporally aligned music videos. SymMV is a large-scale dataset curated from YouTube, comprising 1,181 video-music pairs across more than 10 genres, with a total duration of 78.9 hours. 

We preprocess all datasets by removing vocals and normalizing audio loudness. Silent segments longer than 3 seconds are discarded. The remaining data is segmented into clips up to 30 seconds with a 10-second hop size. Both datasets are split into training, validation, and test sets with an 8:1:1 ratio. In addition, we include V2M-Bench \cite{10} as an out-of-domain test set to evaluate the generalization performance. V2M-Bench contains 300 video-music pairs with a total duration of 9 hours, covering a diverse range of genres including movie trailers, ads, documentaries, and vlogs. The statistics of the processed datasets are shown in Table \ref{tab1}. 
\begin{table}[t]
	\centering
    \resizebox{.7\linewidth}{!}{%
		\begin{threeparttable}
			\begin{tabular}{lllll}
				\toprule
				Dataset& Training & Validation &Test \\
				\midrule
				  BGM909 &8510 &1074 &1061 \\
                    SymMV &9898 &1260 &1245 \\
                    V2M-Bench &0 &0 &1426 \\
				\bottomrule
			\end{tabular}
		\end{threeparttable}
    }
	\caption{The statistical distribution of the adopted datasets}
	\label{tab1}
\end{table}
\subsection{Implementation Details}
Video frames and audio are sampled at 1 FPS and 44.1 kHz, respectively. Rhythmic features have shape $[M, d]$, where $M$ denotes the video duration in seconds, $d = 16$ when using low-resolution Mel-spectrograms or tempograms, and $d = 1$ for low-resolution ODF. Diff-V2M uses a frozen VAE from Stable Audio Open \cite{34}, while the DiT-based diffusion model is trained from scratch. The DiT is optimized with the v-objective \cite{47} to predict noise. During inference, we use a 250-step DDIM sampler with classifier-free guidance (scale 3.0). Training employs AdamW optimizer with a learning rate of $1 \times 10^{-4}$, betas $(0.9,\ 0.999)$, weight decay $1 \times 10^{-3}$, and an InverseLR scheduler (power 0.5). All models are trained for 50 epochs on 2 NVIDIA A100 GPUs.
\subsection{Evaluation Metrics} \label{metrics}
Following the state-of-the-art method \cite{10}, we quantitatively evaluate our model using several metrics that assess the fidelity and diversity of the generated music, including \textbf{Frechet Audio Distance (FAD)}, \textbf{Frechet Distance (FD), Kullback–Leibler divergence (KL)} \cite{33}, \textbf{Density (Den.)} and \textbf{Coverage (Cov.)} \cite{48}. We also use the \textbf{ImageBind Score (IB)} \cite{49} to assess alignment between video and generated music. For subjective evaluation, we consider four criteria \cite{10}, i.e., (1) \textbf{Audio Quality}: perceptual clarity and fidelity of the audio; (2) \textbf{Musicality}: the aesthetic quality of the music, distinct from audio quality; (3) \textbf{Video-Music Alignment}: how well the music matches the visuals; and (4) \textbf{Overall Assessment}: overall generation quality.
\subsection{Comparison Models}
We compare Diff-V2M with the following methods:
\begin{itemize}[leftmargin=0pt]
\item \textbf{CMT} \cite{8} establishes the rhythmic relationships between video
and background music, then proposes a Controllable Music Transformer (CMT) for local rhythmic and global genre/instrument control.

\item \textbf{Video2Music} \cite{15} extracts semantic, scene offset, motion, and emotion features from music videos and proposes an Affective Multimodal Transformer (AMT) to generate music given a video.

\item \textbf{MuMu-LLaMA} \cite{39} combines ViViT and LLaMA \cite{50} with multimodal adapters, then projects audio tokens from LLaMA as conditions for text-to-music generation.

\item \textbf{GVMGen} \cite{11} encodes audio into discrete tokens using EnCodec \cite{51} and predicts discrete audio tokens with hierarchical spatial and temporal cross-attention to align visual features with music. 

\item \textbf{VidMuse} \cite{10} predicts the discrete audio tokens by incorporating local and global visual cues, and employs long-short-term modeling to ensure coherence between the video and music.
\end{itemize}
Note that the first two approaches generate symbolic music, while the others directly generate musical audio.
\subsection{Experimental Results}
\begin{table}[t]
	\centering
	\resizebox{1\linewidth}{!}{%
		\begin{threeparttable}
			\begin{tabular}{lllllll}
				\toprule
				\multirow{2}{*}{Models} & \multicolumn{6}{c}{Metrics}             \\ \cline{2-7} 
                    & FAD$\downarrow$ & FD$\downarrow$ & KL$\downarrow$ & Den.$\uparrow$ & Cov.$\uparrow$ & IB$\uparrow$ \\
				\hline
				\multicolumn{7}{c}{\cellcolor[HTML]{EFEFEF}Mixed Test Set}   \\
                    \hline
                    $\text{Mel}_{LR}$ &2.0376	&13.1304	&\textbf{0.9341}	&0.3690	&\textbf{0.3380}	&0.1653\\
                    $\text{Tem}_{LR}$ &2.3163	&13.0197	&0.9551	&0.3210	&0.3270	&0.1636 \\
                    $\text{ODF}_{LR}$  &\textbf{2.0175}	&\textbf{12.8566}	&0.9432	&\textbf{0.3694}	&0.3370 &\textbf{0.1831} \\
                    \hline
                    \multicolumn{7}{c}{\cellcolor[HTML]{EFEFEF}V2M-Bench}              \\
                    \hline
                    $\text{Mel}_{LR}$ &2.0310  &24.8375  &1.2638  &\textbf{0.6916} &0.3822 &0.1810  \\
                    $\text{Tem}_{LR}$ &2.0230  &\textbf{20.2481}  &1.2939  &0.5170 &\textbf{0.4137} &0.1748   \\
                    $\text{ODF}_{LR}$  &\textbf{1.8129}  &21.3321  &\textbf{1.2405}  &0.6360 &0.3717 &\textbf{0.1887}   \\
				\bottomrule
			\end{tabular}
		\end{threeparttable}
	}
	\caption{The comparison of different rhythmic representations. $\text{Mel}_{LR}$, $\text{Tem}_{LR}$, and $\text{ODF}_{LR}$ refer to low-resolution Mel-spectrogram, tempogram and ODF, respectively. The best results are highlighted in \textbf{bold}.}
	\label{tab2}
\end{table}
\subsubsection{Comparison of Rhythmic Representations.}
Table \ref{tab2} compares the performance of different rhythmic representations, evaluated on both the mixed test set of two datasets and the V2M-Bench dataset. The low-resolution ODF ($\text{ODF}_{LR}$), offering a simpler and more direct rhythmic representation, consistently outperforms other representations. Low-resolution Mel-spectrogram ($\text{Mel}_{LR}$), despite prior use in video-to-sound effect task \cite{19}, is less effective here. Thus, $\text{ODF}_{LR}$ is used as the default rhythmic representation in subsequent experiments. Note that simple additive fusion is employed in this comparison.
\begin{table}[t]
	\centering
	\resizebox{1\linewidth}{!}{%
		\begin{threeparttable}
			\begin{tabular}{lllllll}
				\toprule
				\multirow{2}{*}{Strategies} & \multicolumn{6}{c}{Metrics}             \\ \cline{2-7} 
                    & FAD$\downarrow$ & FD$\downarrow$ & KL$\downarrow$ & Den.$\uparrow$ & Cov.$\uparrow$ & IB$\uparrow$ \\
				\hline
                    Weighted Fusion &2.3625 &11.5332 &0.9246	&0.3106	&0.3540 &0.1729    \\
                    Additive Fusion &2.0175	&12.8566	&0.9432	&0.3694	&0.3370 &\textbf{0.1831}    \\
                    Feature Selection &2.0894 &12.4990	&0.8875	&0.3228	&0.3720 &0.1800    \\ \hline
                    PreAttn FiLM  &2.0812 &13.9032 &0.9507 &0.3192 &0.3500 &0.1640    \\
                    PostAttn FiLM  &2.1340 &12.1286	&0.9052	&0.3682	&0.3510 &0.1808 \\
                    \qquad\qquad w/ FS  &\textbf{1.5175} &\textbf{10.9567} &\textbf{0.8575} &\textbf{0.3756} &\textbf{0.3990} &0.1812    \\
				\bottomrule
			\end{tabular}
		\end{threeparttable}
	}
	\caption{The comparison of different feature fusion strategies on the mixed test set. FS denotes feature selection.}
	\label{tab3}
\end{table}
\subsubsection{Comparison of Feature Fusion Strategies.}
The comparison results of the fusion strategies for the semantic and rhythmic features within the hierarchical conditioning module are presented in Table \ref{tab3}. Note that the weighted fusion following FiLM employs a simple additive fusion strategy. Among post-attention strategies, both feature selection and post-attention FiLM outperform weighted fusion and perform on par with additive fusion. Pre-attention FiLM is less effective than its post-attention counterpart. The best performance is achieved by combining post-attention FiLM with feature selection. 
\begin{table}[t]
	\centering
	\resizebox{1\linewidth}{!}{%
		\begin{threeparttable}
			\begin{tabular}{lllllll}
				\toprule
				\multirow{2}{*}{Models} & \multicolumn{6}{c}{Metrics}             \\ \cline{2-7} 
                    & FAD$\downarrow$ & FD$\downarrow$ & KL$\downarrow$ & Den.$\uparrow$ & Cov.$\uparrow$ & IB$\uparrow$ \\
				\hline
				\multicolumn{7}{c}{\cellcolor[HTML]{EFEFEF}Mixed Test Set}   \\
                    \hline
                    GT           &0.0000  &0.0000   &0.0000 &1.1178 &0.9980 &0.2154 \\
                    CMT          &8.9265 &47.7599  &1.0984 &0.0422 &0.0080 &0.0820  \\
                    Video2Music  &31.1963 &106.7122 &1.7220 &0.0010 &0.0010 &0.0312  \\
                    MuMu-LLaMA   &\underline{2.8410} &27.1160  &1.2481 &0.1074 &0.0900 &0.1448  \\
                    GVMGen       &4.3354 &27.9350  &1.1633 &0.0900 &0.0830 &0.1760  \\
                    VidMuse      &3.4376 &\underline{21.0391}  &\underline{0.9361} &\underline{0.1496} &\underline{0.1300} &\underline{0.1804}  \\
                    Diff-V2M (ours)     &\textbf{1.5175} &\textbf{10.9567} &\textbf{0.8575} &\textbf{0.3756} &\textbf{0.3990} &\textbf{0.1812}   \\\hline
                    \multicolumn{7}{c}{\cellcolor[HTML]{EFEFEF}V2M-Bench}              \\
                    \hline
                    GT          &0.0000	 &0.0000   &0.0000 &1.0632 &0.9930 &0.2379 \\
                    CMT         &10.9118 &65.5007  &1.4109 &0.0895 &0.0189 &0.1255 \\
                    Video2Music &32.4392 &128.9256 &2.1149 &0.0094 &0.0007 &0.0388 \\
                    MuMu-LLaMA  &3.8593	 &40.4072  &1.4866 &0.3637 &0.1971 &0.1714 \\
                    GVMGen      &\underline{2.1528}	 &\textbf{21.5488}  &\underline{1.2060} &\underline{0.3853} &\underline{0.2938} &\textbf{0.2030} \\
                    VidMuse     &2.5875	 &22.0282  &\textbf{1.0138} &0.2181 &0.2020 &0.1963 \\
                    Diff-V2M (ours)  &\textbf{1.7612}  &\underline{22.0164}  &1.2684 &\textbf{0.5083} &\textbf{0.4032} &\underline{0.1967}    \\
				\bottomrule
			\end{tabular}
		\end{threeparttable}
	}
	\caption{Comparison with existing methods. The best results are highlighted in \textbf{bold}, and the second-best are \underline{underlined}.}
	\label{tab4}
\end{table}
\begin{figure}[t]
\centering
\includegraphics[width=.97\linewidth]{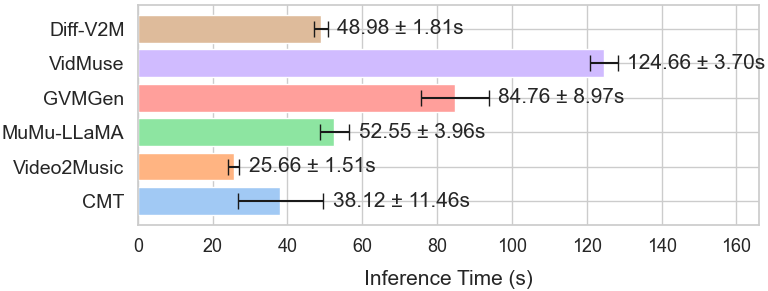} 
\caption{The comparison of inference time for different methods in generating soundtracks for 30-second videos.}
\label{fig4}
\end{figure}
\subsubsection{Quantitative Comparisons with Other Methods.}
Quantitative results in Table~\ref{tab4} show that the proposed Diff-V2M significantly outperforms prior methods on the in-domain test sets. On the out-of-domain V2M-Bench dataset, Diff-V2M also achieves the best overall performance, particularly in audio quality. However, it lags slightly behind GVMGen in video-music alignment, likely because GVMGen was trained on larger, more diverse video datasets similar to V2M-Bench. Additionally, Figure~\ref{fig4} reports average inference time over 10 runs for generating 30-second music clips. CMT and Video2Music are faster as they generate simpler symbolic music. While audio-based models are slower, Diff-V2M achieves the shortest inference time among them.
\begin{figure}
\centering
\includegraphics[width=.95\linewidth]{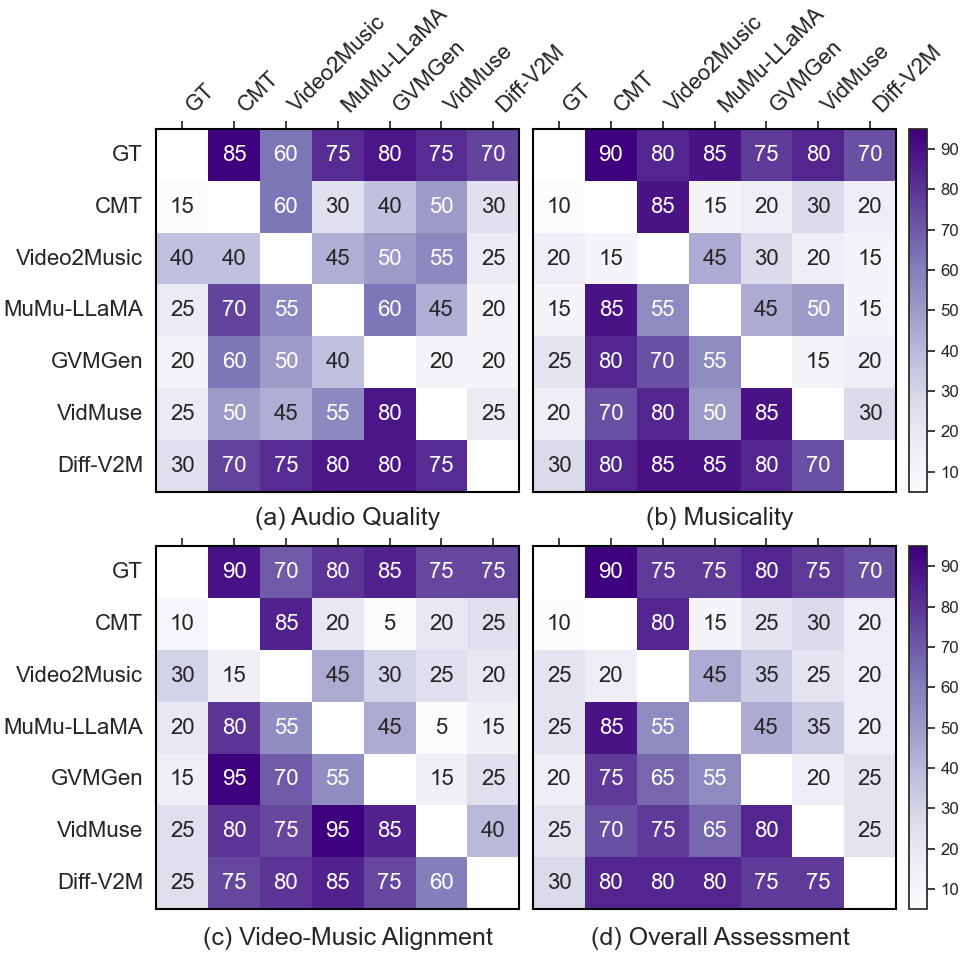} 
\caption{A/B test results of the subjective comparisons.}
\label{fig5}
\end{figure}
\subsubsection{Subjective Evaluation.}
We conducted an A/B test to subjectively compare different methods. A total of 30 participants (20 amateurs and 10 experts) were invited to ensure that each pair of methods was compared 20 times. Participants were instructed to choose their preferred sample based on the subjective criteria. The evaluation results are presented in Figure~\ref{fig5}, where the value at position $\text{matrix}[i][j]$ indicates the percentage (0–100) that the method in row $i$ was preferred over that in column $j$. Diff-V2M outperforms all baselines in over half of the pairwise comparisons, except when compared with ground-truth (GT) samples. These results highlight the subjective superiority of Diff-V2M.
\begin{table}[t]
	\centering
	\resizebox{1\linewidth}{!}{%
		\begin{threeparttable}
			\begin{tabular}{lllllll}
				\toprule
				\multirow{2}{*}{Ablation} & \multicolumn{6}{c}{Metrics}             \\ \cline{2-7} 
                    & FAD$\downarrow$ & FD$\downarrow$ & KL$\downarrow$ & Den.$\uparrow$ & Cov.$\uparrow$ & IB$\uparrow$ \\
				\hline
                    Diff-V2M &\textbf{1.5175} &10.9567 &\textbf{0.8575} &\underline{0.3756} &\textbf{0.3990} &0.1812    \\
                    w/o $C_\mathbf{r}$ &1.8264 &11.9535 &\underline{0.8692} &0.3476 &0.3610  &\textbf{0.1893}    \\
                    w/o $C_\mathbf{e}$  &1.6761 &12.8922 &0.9306 &0.3664 &0.3480 &0.1814    \\
                    w/o $C_\mathbf{r}$ \& $C_\mathbf{e}$ &1.7086 &\textbf{9.7514} &0.8775 &0.3636 &\underline{0.3900} &0.1814 \\
                    w/o VR  &2.2177 &13.6060 &0.9147 &0.3220 &0.3300 &0.1800  \\ 
                    w/o Joint &1.8787 &13.3894 &0.8743 &\textbf{0.3798} &0.3880 &0.1814  \\ 
                    w/o Scheduler &\underline{1.6159} &\underline{10.6706} &0.9105 &0.3402 &0.3630 &\underline{0.1860}  \\ 
				\bottomrule
			\end{tabular}
		\end{threeparttable}
	}
	\caption{Ablation results of different training strategies for Diff-V2M on the mixed test set. }
	\label{tab5}
\end{table}
\subsubsection{Ablation Studies.}
We further conduct ablation studies on Diff-V2M’s training strategies: (1) w/o $C_\mathbf{r}$, remove rhythmic features; (2) w/o $C_\mathbf{e}$, remove emotional features; (3) w/o $C_\mathbf{r}$ \& $C_\mathbf{e}$: remove both; (4) w/o Visual Rhythm (VR): the rhythm predictor takes only CLIP features as input, excluding dynamic visual inputs including scene onset and visual beat; (5) w/o Joint: train the rhythm predictor and music generator separately, leading to a training–inference mismatch; (6) w/o Scheduler: jointly train the predictor and generator from the beginning without applying scheduled conditioning. Table~\ref{tab5} shows that Diff-V2M achieves the best performance by incorporating rhythmic and emotional features as well as adopting the scheduled training strategy. Removing VR degrades performance, likely due to reduced rhythm prediction accuracy. Interestingly, excluding $C_\text{r}$ slightly improves ImageBind (IB) score, possibly because IB emphasizes semantic alignment, which may benefit from the absence of rhythm-related interference during generation. 
\section{Conclusion}
We propose Diff-V2M, a general video-to-music generation framework designed to handle diverse videos. To explicitly model musical rhythm, we introduce a simple yet effective rhythmic representation and develop a predictor to estimate it from video. Diff-V2M builds upon an audio LDM, featuring a hierarchical cross-attention conditioning module to integrate multiple video-derived features, along with novel fusion strategies to adaptively combine semantic and rhythmic cues. Extensive comparisons on both mixed in-domain and out-of-domain test sets demonstrate the superiority of Diff-V2M in both quantitative and qualitative evaluations.
\paragraph{Limitation.} Despite promising performance, our method has limitations. First, relying on scene cuts and inter-frame differences may overlook subtle motion cues, leading to suboptimal rhythm alignment in human-centric videos. Second, the model lacks explicit control over musical attributes such as genre and emotion, limiting its adaptability in scenarios requiring style or affective manipulation. These limitations highlight valuable directions for future research.

\section{Acknowledgments}
This work was supported by the National Natural Science Foundation of China (No.62272409).

\bibliography{aaai2026}

\clearpage
\appendix
\twocolumn[
\begin{center}
	\huge \textbf{Appendix}
\end{center}
]
\section{Architecture of Diff-V2M}
The training flow of Diff-V2M is shown in Algorithm~\ref{al1}.
\begin{algorithm}
	\caption{Diff-V2M}
	\label{al1}
	\KwIn{Video clip $V$}
	\KwOut{Generated musical audio $A$}
	
	\textbf{Waveform Encoding:} \\
	Latent representation $z = \text{VAE Encoder}(A)$\\
	\textbf{Visual Feature Extraction:} \\
	Extract semantic features: $C_\mathbf{s} = \text{CLIP}(V)$ \\
	Extract emotional features: $C_\mathbf{e} = \text{ColorHist}(V)$ \\
	Extract visual rhythm: $C_\mathbf{VR} = \text{VisualDynamics}(V)$
	
	\textbf{Rhythm Prediction:} \\
	Predict rhythm features: $C_\mathbf{r} = \text{RhythmPredictor}(C_\mathbf{s}+C_\mathbf{VR})$ \\
	Apply scheduled conditioning: \\
	\Indp 
	Sample $q \sim \text{Bernoulli}(p_\text{pred}(e))$ at epoch $e$ \\
	\If{$q = 1$}{
		use $C_\mathbf{r}$
	}
	\Else{
		use ground-truth rhythm $C_\mathbf{r}^{gt}$
	}
	\Indm
	
	\textbf{Hierarchical Conditional Diffusion:} \\
	Add noise to encoded latent: $z^t = \text{AddNoise}(z)$ \\
	\For{timestep $t = 1$ to $0$}{
		
		\textbf{Global Conditioning:} \\
		Concatenate global metadata: $C_\mathbf{g} = [g_\text{start}; g_\text{dur}]$ \\
		Compute time-modulated global embedding: $C_\mathbf{g}^t = \text{MLP}([C_\mathbf{g}+\text{Embed}(t)])$ \\
		Prepend $C_\mathbf{g}^t$ to $z$ and apply RoPE \\
		$\tilde{z} = \text{RoPE}([C_\mathbf{g}^t; z^t])$ \\
		\For{\text{Block} $l = 1$ to $N$}{
			\textbf{Cross Attention 1} (emotional condition) \\
			\Indp $\mathbf{h}_{\text{emo}}^{(l)} \leftarrow \text{CrossAttn}(\mathbf{h}_{\text{self}}^{(l)}, \mathbf{K}_\mathbf{e},\mathbf{V}_\mathbf{e})$ \\
			\Indm
			\textbf{Cross Attention 2} (semantic and rhythmic conditions) \\
			\Indp
			$\mathbf{h}_{\text{sem}}^{(l)} = \text{CrossAttn}(\mathbf{h}_{\text{emo}}^{(l)}, \mathbf{K}_\mathbf{s},\mathbf{V}_\mathbf{s})$ \\
			$\mathbf{h}_{\text{rhy}}^{(l)} = \text{CrossAttn}(\mathbf{h}_{\text{emo}}^{(l)}, \mathbf{K}_\mathbf{r},\mathbf{V}_\mathbf{r})$ \\
			\Indm
			\textbf{Feature fusion followed by FFN} \\
			\Indp
			$\mathbf{h}^{(l+1)} \leftarrow \text{FFN}(\text{Fuse}(\mathbf{h}_{\text{sem}}^{(l)}, \mathbf{h}_{\text{rhy}}^{(l)}, t))$ \\
			\Indm
		}
	}
	Denoise latent: $z \leftarrow \text{DiT}_\theta(z^t, \mathbf{C}, t)$
	
	\textbf{Waveform Reconstruction:} \\
	Generate audio $A = \text{VAE Decoder}(z)$\\
	\Return $A$
\end{algorithm}
\subsubsection{DiT-based Conditional Music Generator}
In this paper, conditions \(\mathbf{C}\) include features extracted by specialized encoders, i.e., emotional, semantic, and rhythmic features, along with global embeddings including music start time $g_\text{start}$ and duration $g_\text{dur}$. These two embeddings are concatenated into a single vector and added to a learnable timestep embedding corresponding to the diffusion step $t$. The time-modulated global embedding is then prepended as a special token to the input sequence. To enhance temporal awareness within the attention mechanism, we further adopt rotary positional embeddings (RoPE), dynamically generating the position encoding matrix based on the current sequence length. The enhanced sequence representation is subsequently fed into the diffusion transformer.

\subsubsection{Hierarchical Cross-attention Module.}
Let \( \mathbf{h}^{(l)} \) be the input sequence at the \( l \)-th Transformer block. After the self-attention module, the emotional context is first integrated as:
\begin{equation}
	\begin{aligned}
		\mathbf{h}_\text{self}^{(l)}&=\text{SelfAttn}(\mathbf{h}^{(l)})\\
		\mathbf{h}_{\text{emo}}^{(l)} &= \text{CrossAttn}(\mathbf{h}_\text{self}^{(l)}, \mathbf{K}_e, \mathbf{V}_e).
	\end{aligned}
\end{equation}
where $\mathbf{K}_e$ and $\mathbf{V}_e$ are obtained through linear projections of emotional features $C_\mathbf{e}$. Using $\mathbf{h}_{\text{emo}}^{(l)}$ as the updated query, we compute cross attention with semantic and rhythmic features in parallel:
\begin{equation}
	\begin{aligned}
		\mathbf{h}_{\text{sem}}^{(l)} &= \text{CrossAttn}(\mathbf{h}_{\text{emo}}^{(l)}, \mathbf{K}_s, \mathbf{V}_s), \\
		\mathbf{h}_{\text{rhy}}^{(l)} &= \text{CrossAttn}(\mathbf{h}_{\text{emo}}^{(l)}, \mathbf{K}_r, \mathbf{V}_r).
	\end{aligned}
\end{equation}
where $\mathbf{K}_s$ and $\mathbf{V}_s$, and $\mathbf{K}_r$ and $\mathbf{V}_r$, are obtained through linear projections of the semantic features $C_\mathbf{s}$ and the rhythmic features $C_\mathbf{r}$, respectively. Timestep-aware feature fusion modules then adaptively combine both branches before the final feed-forward network (FFN), yielding the  $(l{+}1)$-th input sequence: 
\begin{equation}
	\mathbf{h}^{(l+1)} = \text{FFN}\left(\text{Fuse}(\mathbf{h}_{\text{sem}}^{(l)}, \mathbf{h}_{\text{rhy}}^{(l)},t)\right).
\end{equation}
\setcounter{table}{5}
\begin{table*}[t]
	\centering
		\begin{threeparttable}
			\begin{tabular}{lllllll}
				\toprule
				\multirow{2}{*}{Strategies} & \multicolumn{6}{c}{Metrics}             \\ \cline{2-7} 
				& FAD$\downarrow$ & FD$\downarrow$ & KL$\downarrow$ & Den.$\uparrow$ & Cov.$\uparrow$ & IB$\uparrow$ \\
				\hline
				Weighted Fusion &2.3050	&\textbf{18.5891} &1.3059 &0.5196 &\textbf{0.4299}	&0.1764 \\
				Additive Fusion &1.8129	&21.3321 &\textbf{1.2405} &0.6360	&0.3719	&0.1887 \\
				Feature Selection &1.9527 &25.7266 &1.2601 &0.5149	&0.3864	&0.1867 \\ \hline
				PreAttn FiLM  &2.3738 &23.7385 &1.2626 &\textbf{0.7844} &0.3941	&0.1703 \\
				PostAttn FiLM &1.9276 &20.6385 &1.2583 &0.6196 &0.3717	&0.1923 \\
				\qquad\qquad w/ FS  &\textbf{1.7612}  &22.0164  &1.2684 &0.5083 &0.4032 &\textbf{0.1967}    \\
				\bottomrule
			\end{tabular}
		\end{threeparttable}
	\caption{The comparison of different feature fusion strategies on the V2M-Bench dataset. FS denotes feature selection. The best results are highlighted in \textbf{bold}.}
	\label{tab6}
\end{table*}
\begin{table*}[t]
	\centering
		\begin{threeparttable}
			\begin{tabular}{lllllll}
				\toprule
				\multirow{2}{*}{Ablation} & \multicolumn{6}{c}{Metrics}             \\ \cline{2-7} 
				& FAD$\downarrow$ & FD$\downarrow$ & KL$\downarrow$ & Den.$\uparrow$ & Cov.$\uparrow$ & IB$\uparrow$ \\
				\hline
				Diff-V2M &\textbf{1.7612}  &22.0164  &1.2684 &0.5083 &\textbf{0.4032} &\textbf{0.1967}     \\
				w/o $C_\mathbf{r}$ &\underline{1.7671} &22.5636	&1.2595	&0.5613	&0.3822	&0.1913 \\
				w/o $C_\mathbf{e}$  &1.9417	&24.5074 &1.3307 &0.4659	&0.3576	&\underline{0.1959} \\
				w/o $C_\mathbf{r}$ \& $C_\mathbf{e}$ &1.7797 &\underline{21.2935}	&\underline{1.2345}	&\textbf{0.6144}	&\underline{0.4004}	&0.1906 \\
				w/o VR  &2.0195	&22.4100 &1.2864 &0.4300 &0.3892	&0.1861 \\ 
				w/o Joint &3.3245 &30.2149 &\textbf{1.2305} &\underline{0.5798} &0.2854	&0.1753 \\ 
				w/o Scheduler &1.7706 &\textbf{19.9180} &1.3037 &0.4861 &0.3850	&0.1910 \\ 
				\bottomrule
			\end{tabular}
		\end{threeparttable}
	\caption{Ablation results of different training strategies for Diff-V2M on V2M-Bench dataset. The best results are highlighted in \textbf{bold}, and the second-best are \underline{underlined}.}
	\label{tab7}
\end{table*}
\subsubsection{Feature Selection}
According to Diff-BGM \cite{37}, models tend to generate the melody, which is influenced by the semantics, and then generate the rhythm of the music, which is related to the dynamic feature of the video. Based on this, a feature selector is devised to enforce exclusive attention to a single feature at each time step, neglecting the complementary information from other features. This rigid selection limits the model’s ability to leverage multiple features in complex video scenarios, especially when both semantic and rhythmic information are crucial. 

Diff-BGM selects conditioning features during the denoising process from timestep \(N\) to \(0\) based on a hyper-parameter \(t_0\). Specifically, language features are used when \(t_0 > 200\), and video features when \(t_0 \leq 200\). In this paper, we follow a similar strategy. Since our model adopts a continuous noise schedule with timesteps normalized to $t\in[0,1]$, we select semantic features when \(t_0 > 0.2\), and rhythmic features when \(t_0 \leq 0.2\).
\subsubsection{Algorithm.} The overall training workflow of Diff-V2M is presented in \textbf{Algorithm 1}. Note that the self-attention processes within the transformer block are omitted, as follows: 
\begin{equation}
	\mathbf{h}_\text{self}^{(l)} =
	\left\{
	\begin{aligned}
		&\text{SelfAttn}(\tilde{z}), && \text{if } l = 1 \\
		&\text{SelfAttn}(\mathbf{h}^{(l)}), && \text{if } l > 1
	\end{aligned}
	\right.
\end{equation}
\section{Experiments}
\subsubsection{Comparison of Inference Time.}
We compare the average inference time over 10 runs for generating 30-second music clips using different methods. All models were evaluated on a single A10 GPU (20 GB), except for MuMu-LLaMA, which was executed on an A100 GPU (80 GB) due to its higher memory requirements. Note that GVMGen is limited to generating background music for videos of up to 25 seconds in duration.
\subsubsection{Comparison of Feature Fusion Strategies.}
Table~\ref{tab6} presents a comparison of feature fusion strategies on the V2M-Bench dataset. Among them, post-attention FiLM combined with feature selection achieves the best overall performance in terms of audio quality and video-music alignment, while weighted fusion yields higher diversity in generated music. Since V2M-Bench serves as an out-of-domain dataset, the best scores are distributed across different strategies rather than concentrated in a model.
\subsubsection{Ablation Studies.}
Ablation study results on the V2M-Bench dataset are shown in Table~\ref{tab7}. Consistent with the findings on the in-domain mixed test sets, Diff-V2M achieves the best performance by incorporating rhythmic and emotional features and leveraging the scheduled conditioning training strategy.

\end{document}